\documentclass[aps,prb,twocolumn,groupedaddress,noshowpacs,checklength,amsmath]{revtex4-1}

\usepackage{graphicx}
\usepackage{amssymb}
\usepackage{amsmath}
\usepackage{cancel}
\usepackage{color}

\newcommand{\grad}{\mbox{\boldmath $\nabla$}}

\def\xhat{{\bf \hat x}}
\def\yhat{{\bf \hat y}}
\def\zhat{{\bf \hat z}}

\def\phat{{\hat{\mbox{\boldmath$\phi$}}}}

\def\nhat{{\bf \hat n}}
\def\that{{\bf \hat t}}

\def\bhat{{\bf \hat b}}
\def\shat{{\bf \hat s}}

\def\rcurs{{\mbox{$\resizebox{.16in}{.08in}{\includegraphics{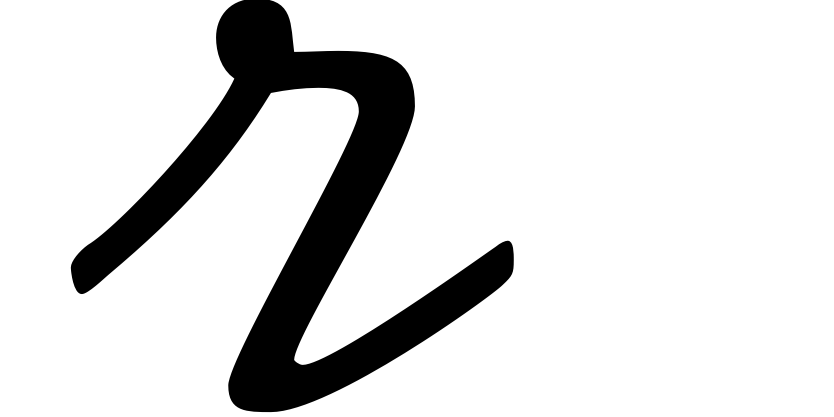}}$}}}
\def\brcurs{{\mbox{$\resizebox{.16in}{.08in}{\includegraphics{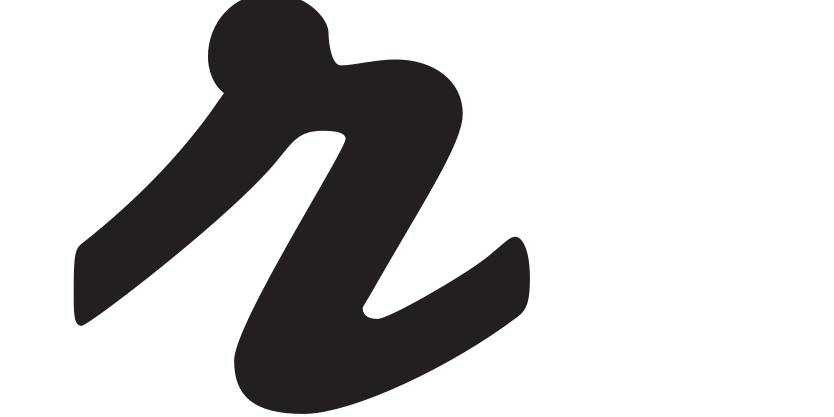}}$}}}

\begin{document}

\title{A Taxonomy of Magnetostatic Field Lines}

\author{Joel Franklin}\email[Electronic address: ]{jfrankli@reed.edu}
\author{David Griffiths}\email[Electronic address: ]{griffith@reed.edu}
\author{Darrell Schroeter}\email[Electronic address: ]{schroetd@reed.edu}
\affiliation{Department of Physics,
Reed College, Portland, Oregon  97202}

\begin{abstract}  
Contrary to widespread belief, magnetostatic field lines do not ordinarily form closed loops.  Why, then, are they in fact closed for so many familiar examples?  What other topologies are possible, and what current configurations generate them?
\end{abstract}

\maketitle

\section{Introduction} 
In classical electrodynamics the fundamental physical quantities are electric and magnetic {\it fields}, ${\bf E}({\bf r},t)$ and ${\bf B}({\bf r},t)$.  Field {\it lines},\cite{AFL} which connect up the field vectors at adjacent points, can be useful for visualizing the structure of the fields, but they play no role at all in the underlying theory.  As Feynman put it, ``[F]ield lines \ldots are only a crude way of describing a field and it is very difficult to give the correct, quantitative laws directly in terms of field lines.  Also, the ideas of the field lines do not contain the deepest principle of electrodynamics, which is the superposition principle.  Even though we know how the field lines look for one set of charges and what the field lines look like for another set of charges, we don't get any idea about what the field line patterns will look like when both sets are present together.  [In terms of the fields {\bf E} and {\bf B}], on the other hand, superposition is easy---we simply add the two vectors.''\cite{SIME}  

Moreover, field lines are subject to some common misconceptions.  Introductory students are often told that
\begin{enumerate}
\item Electric field lines cannot terminate in empty space; they originate on positive charges and end on negative charges (or they run off to infinity).
\item Magnetic field lines cannot terminate in empty space; they form closed loops (or they run off to infinity).
\item The density of electric field lines is proportional to the strength of the field.
\item The density of magnetic field lines is proportional to the strength of the field.
\item Field lines are ``real'' physical entities: invisible ``strings'' existing out there in the space around charges and currents.
\end{enumerate}

Of these, only number 3 is true in general\cite{ITIG} (even for {\it static} configurations).  The goal of this paper is to challenge these misapprehensions, by presenting a number of simple explicit counterexamples, and by calling attention to the somewhat scattered literature on the subject.  We do not pretend to have discovered anything fundamentally new, and nothing here will be news to specialists (especially plasma physicists).  But we do suspect that some students (and their instructors) will be surprised to learn that simple rules they were taught in high school are not sustainable.  We show that both electric and magnetic field lines can terminate in mid-air, at points where the field is zero, and we demonstrate that  magnetic field lines do {\it not} ordinarily form closed loops\cite{OFCL} (though this is a striking feature of many familiar examples).  We identify two interesting alternatives, which we call the ``vortex'' (a field line that spirals in to a line current) and the ``slinky'' (a field line that corkscrews along a line current).    

 This raises some surprisingly subtle questions: 
\begin{itemize}
\item What is it about the familiar text-book configurations that leads to closed magnetic field lines?  How can we tell in advance (from the structure of the source current) whether it will generate closed field lines?
\item Often those closed field lines are in fact {\it planar}; what is the connection (if any) between field lines that lie in a plane and field lines that form closed loops?
\item What is the role of special symmetries of the current configuration?
\item Can we classify the topologies of possible non-closed field lines?
\end{itemize}
We cannot provide simple and definitive answers, but this is a start.

In Section II we consider more carefully the notion of a ``field line.''  What exactly does the term mean, and have we perhaps been misled by Faraday and others, who entertained a 
very concrete interpretation that is difficult to justify in Maxwell's electrodynamics?\cite{JIME}  In Section III we explore some surprising examples of magnetic field lines that do {\it not}
 form closed loops.  In Section IV we consider some special symmetries of the source currents, and their implications for field line configurations.  In Section V we return to the question
  of why so many familiar currents lead to closed field lines.  Appendix A derives (from the Biot-Savart law) some results quoted in Section IV.  Appendix B develops an intriguing relation
  between field lines in special two-dimensional problems and contour plots (which are typically closed loops).  And Appendix C discusses the curvature and torsion of field lines, and 
  some implications of the Frenet-Serret formulas in this context.  In the Supplementary materials we offer some useful tools to aid interested readers in exploring configurations of their 
  own devising.

\section{Field Lines}

What, precisely, {\it is} a field line---how, for example would you instruct a computer to plot one, for a specified field ${\bf F}({\bf r})$?  The basic idea is very simple: to get from one point on a field line, ${\bf r}(u)$, to the next, ${\bf r}(u+du)$, you take a short step in the direction of the field: 
\begin{equation}
{\bf r}(u+du) ={\bf r}(u) + \frac{{\bf F}({\bf r}(u))}{|{\bf F}({\bf r}(u))|}\,\lambda\,du.\label{eq1}
\end{equation} 
Here $u$ is any smooth parameter that increases monotonically along the curve, and $\lambda(u)\,du$ is the distance from ${\bf r}(u)$ to ${\bf r}(u+du)$.  In the infinitesimal limit,
\begin{equation}
\frac{d{\bf r}}{du} = \lambda\,\frac{{\bf F}({\bf r})}{|{\bf F}({\bf r})|}. \label{eq2}
\end{equation}
You're free to pick $\lambda > 0$ however you want (it can even be a function of {\bf r}), since $\lambda$ just determines how far along the field line you progress as you increase $u$.  A natural choice is $\lambda = 1$; in that case $u$ is the arc length, $\ell$:
\begin{equation}
\frac{d{\bf r}}{d\ell} = \frac{\bf F({\bf r})}{|{\bf F}({\bf r})|}.\label{eq2.5}
 \end{equation}
 Another convenient possibility---the one we shall use in this paper---is $\lambda = |{\bf F}({\bf r})|$:
 \begin{equation}
 \frac{d{\bf r}}{du} = {\bf F}({\bf r}).\label{eq2.6}
 \end{equation}
In any case, what we have is an ordinary first-order differential equation for ${\bf r}(u)$, and (fourth order) Runge-Kutta is the method of choice for numerical solutions.\cite{BDES}

What happens when a field line approaches a point where the field itself vanishes?\cite{WTFV}  If the field on the ``far side'' has the opposite direction; then the algorithm (\ref{eq1}) will drive the field line back and forth across the null point, effectively terminating it there.\cite{ETIT} For example, imagine two identical positive charges on the $x$ axis, at $x=+a$ and $x=-a$.  A field line departs from the one at $+a$, heading to the left (toward the origin); another departs from the charge at $-a$, heading to the right (also toward the origin).  What happens when the two incompatible field lines collide at the center?  The student is probably told ``just don't draw that field line'' (after all, you can only draw a representative sampling, and at the center, where the field is zero, there shouldn't {\it be} a field line, since the density of field lines is proportional to the strength of the field).  Wouldn't it be better to admit that those two perfectly legitimate field lines simply {\it terminate} at $x=0$, and amend the rule: ``electric field lines can terminate {\it either} at point charges (or at infinity), {\it or else at places where the field vanishes}''?  

The same goes for magnetic fields: Imagine two square wire loops, centered on the $z$ axis, one at $z=b$ and one at $z=-b$, carrying opposite currents.  The field along the $z$ axis points in the $+z$ direction for positive $z$, and in the $-z$ direction for negative $z$ (Fig.~1).  One field line starts at the origin and heads up the positive $z$ axis, while another starts there and heads down the negative $z$ axis.  Meanwhile, if (as we just assumed) that center point is a ``source" for field lines along the $z$ axis, it is a ``sink'' for field lines coming in along the $x$ and $y$ axes: they terminate in midair at the origin, where the field is zero.\cite{TNZA}

\vskip0in
\begin{figure}[h]
\hskip0in\scalebox{1.5}[1.5]{\includegraphics{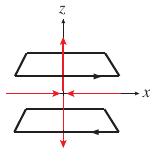}}
\vskip0in
\caption{Field lines that terminate in mid-air (currents black, field lines red).}
\end{figure}

By construction, a field line always ``grows'' in the direction of the field, and if it hits a point where ${\bf F} = {\bf 0}$ (all three components), it stops.  Each field line carries an arrow---its {\it direction}---and it cannot switch its direction in mid-stream.  Except for ``singular'' points, where $|{\bf F}|$ goes to infinity (as, for instance, on a current-carrying wire) or ${\bf F}= {\bf 0}$, there is a (unique) field line through every point in space.  
 
\section{Examples}  
From now on we will restrict our attention to (static) magnetic field lines, which are much more interesting (and problematic) than electric ones.  Every introductory textbook displays the familiar magnetic field lines for long straight wires, circular current loops, solenoids, toroidal coils, infinite current sheets, and magnetic dipoles.  In this Section we will explore some less familiar configurations, to get a feel for the range of possibilities.

\subsection{The Slinky}  
For a circular current, the field lines form plane closed loops around the wire (Fig.~2).\cite{TWFT}  But suppose we now introduce an additional current, along the axis ($z$) of the loop.\cite{AOTL}  Its field lines circle the $z$ axis, and impart an azimuthal component to the total, which now forms a kind of helix, wrapping around the wire loop like a slinky (Fig.~3).  Except for certain special ratios of the two currents the field lines will not close on themselves, but wind around again and again to fill out a toroidal surface.  

\vskip0in
\begin{figure}[h]
\hskip0in\scalebox{.21}[.21]{\includegraphics{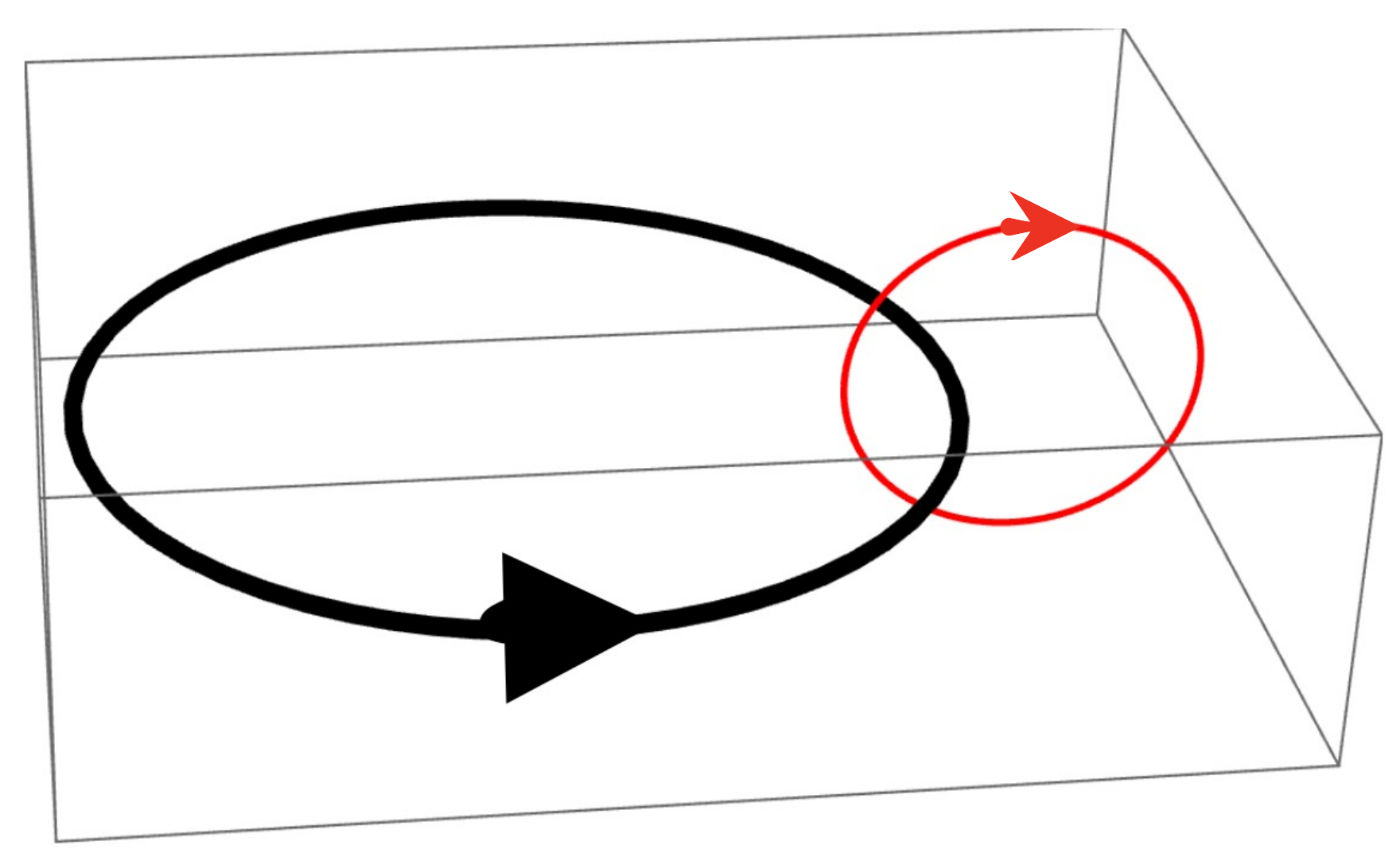}}
\vskip0in
\caption{Field line for a circular current loop.}
\end{figure}

\vskip0in
\begin{figure}[h]
\hskip0in\scalebox{.25}[.25]{\includegraphics{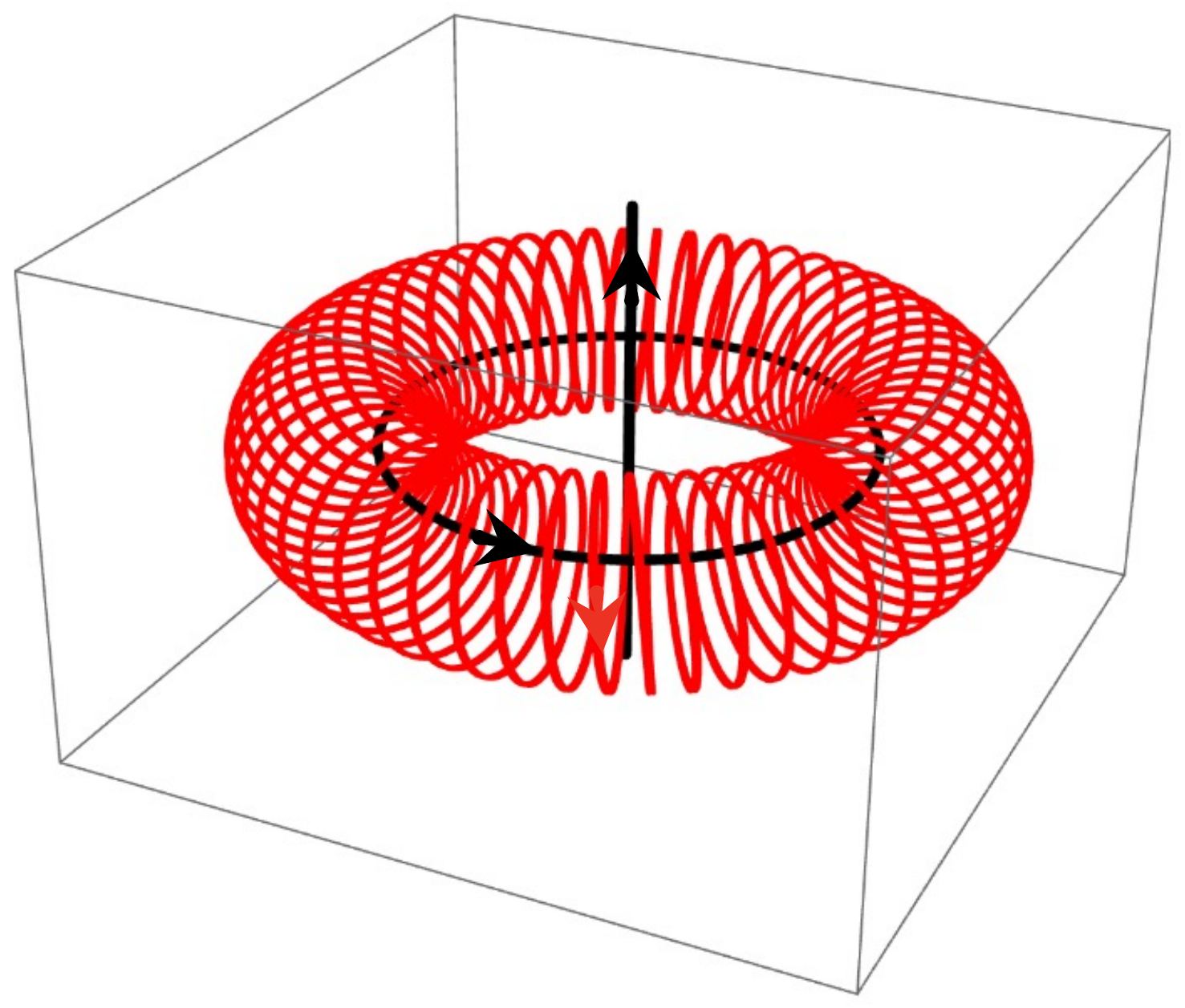}}
\vskip0in
\caption{Slinky on a circular current loop.}
\end{figure}

You can do the same thing for a square current loop (Fig.~4), or even by combining a uniform longitudinal field (say, from an infinite solenoid) and a coaxial line current (Fig.~5)---though in this case the field line does not fill in the whole surface.

\vskip0in
\begin{figure}[h]
\hskip0.3in\scalebox{.25}[.25]{\includegraphics{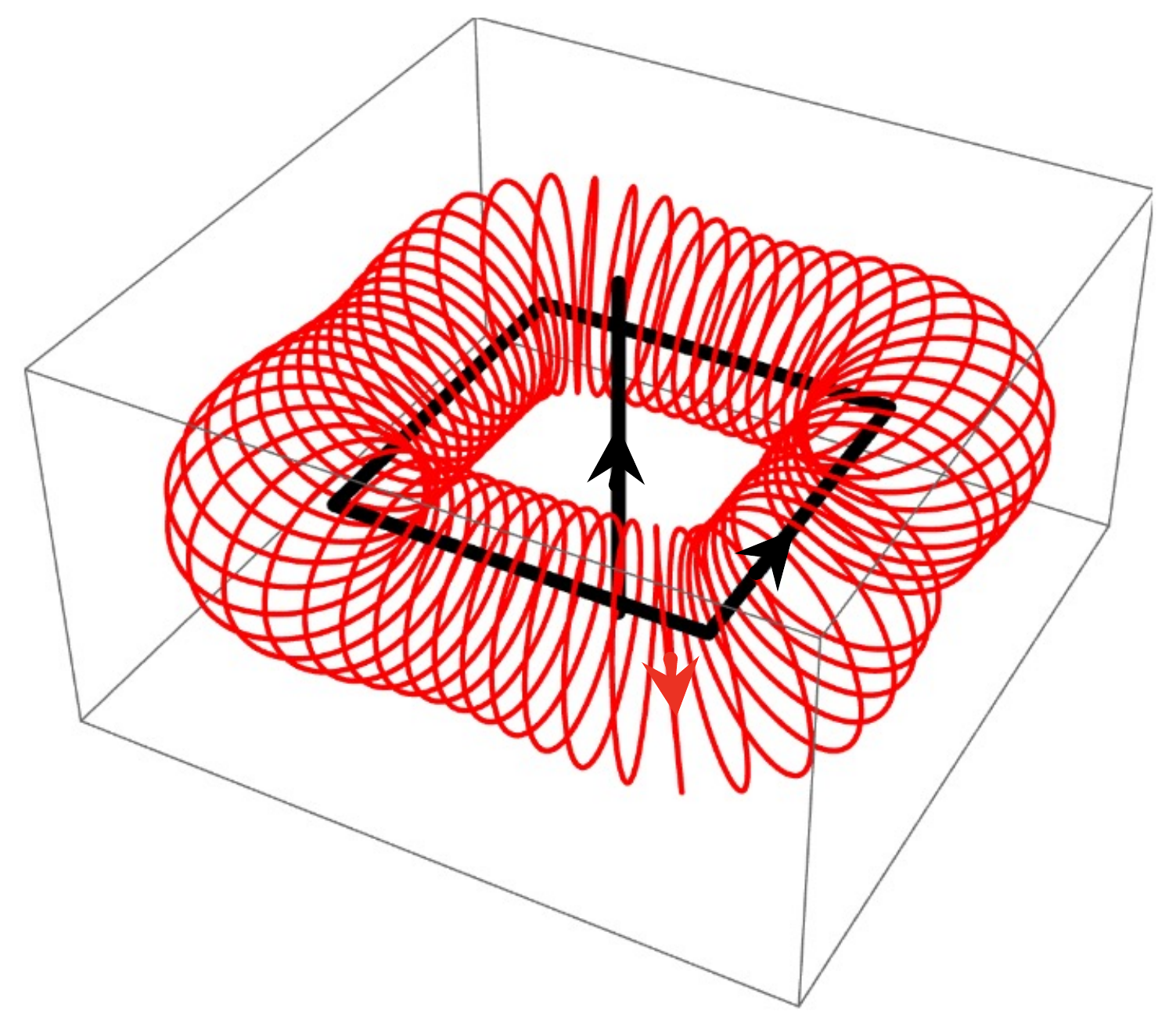}}
\vskip0in
\caption{Slinky on a square current loop.}
\end{figure}

\vskip0in
\begin{figure}[h]
\hskip0in\scalebox{.25}[.25]{\includegraphics{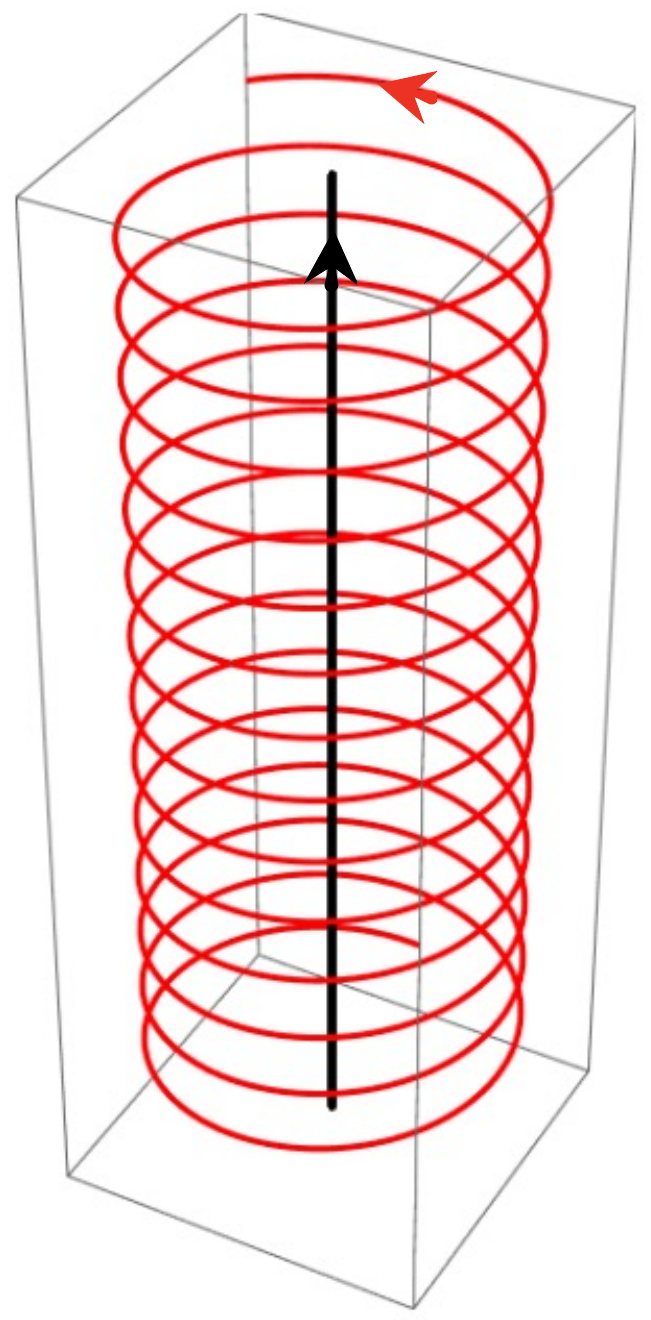}}
\vskip-.1in
\caption{Slinky on an infinite straight current.  (Not shown: the source current for the uniform longitudinal field.)}
\end{figure}

Of course, these configurations are artificial, in the sense that they include wires extending to infinity.  But you can achieve essentially the same effect by replacing the infinite wire with a second circular loop, interlocking the first (Fig.~6).\cite{LITF}

\vskip0in
\begin{figure}[h]
\hskip0in\scalebox{.25}[.25]{\includegraphics{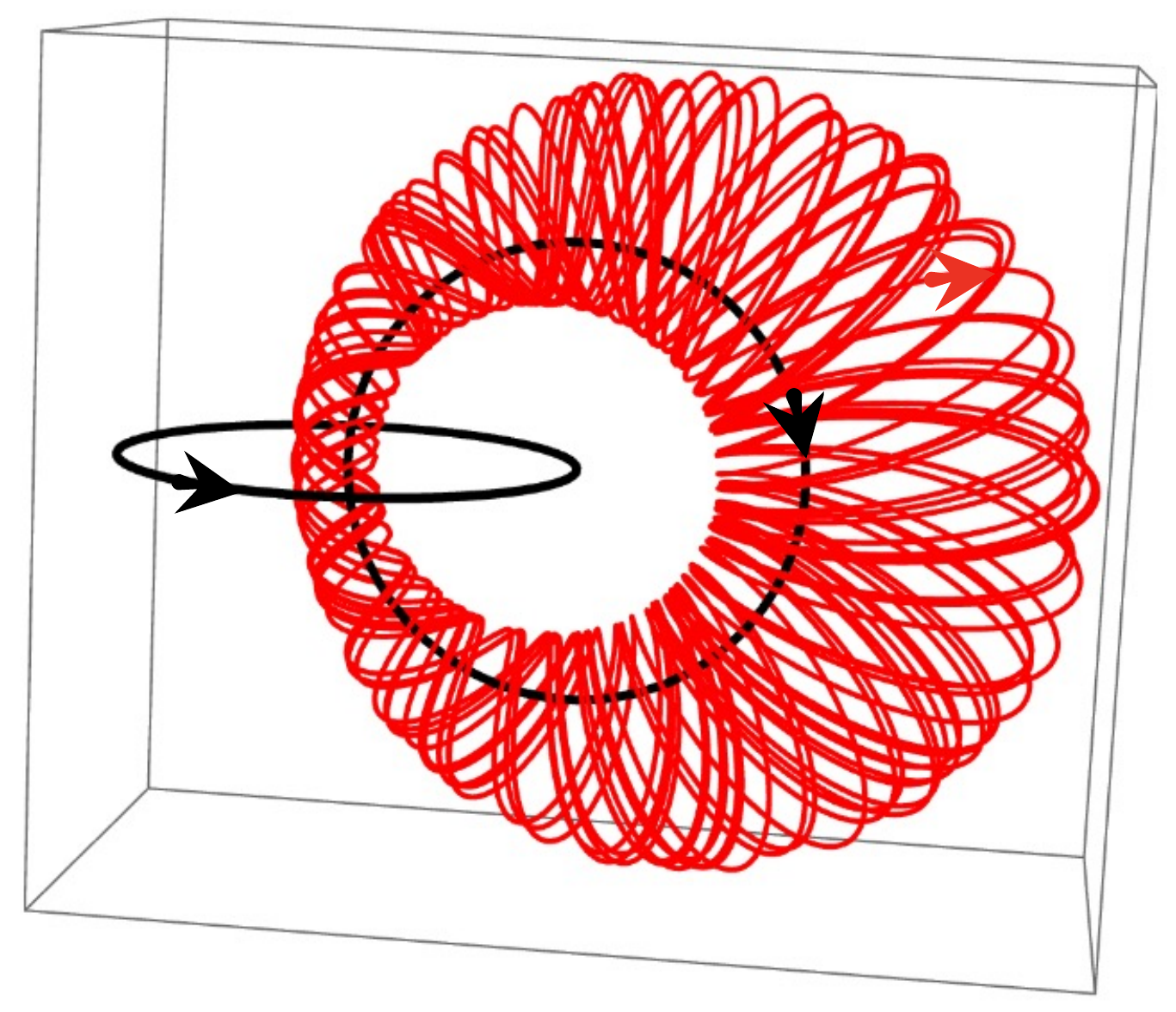}}
\vskip-.1in
\caption{Interlocking circular currents.}
\end{figure}

\subsection{The Vortex} 
Imagine now two parallel square current loops, one directly above the other, and carrying identical currents.  If you don't get too close to either wire, the field lines form closed loops (Fig.~7).  But suppose you start out quite close to (say) the lower square, in the plane bisecting one of its sides?   By symmetry, the field line will remain in that plane.  It ``orbits'' the wire in an expanding spiral, until at some point it is captured by the upper square, and proceeds to spiral inward (Fig.~8).  What if we track the line back in the other direction (by reversing the sign of the current)?  It spirals in, closer and closer to the lower wire, filling in the ``hole'' in the figure.  In both directions, the field line never really terminates, but neither does it form a closed loop; it ``ends'' in a death spiral around the current; we'll call it a ``vortex.''  In this case the field line starts as a vortex on one wire, and ends as a vortex on the other wire.

\vskip0in
\begin{figure}[h]
\hskip0.3in\scalebox{.23}[.23]{\includegraphics{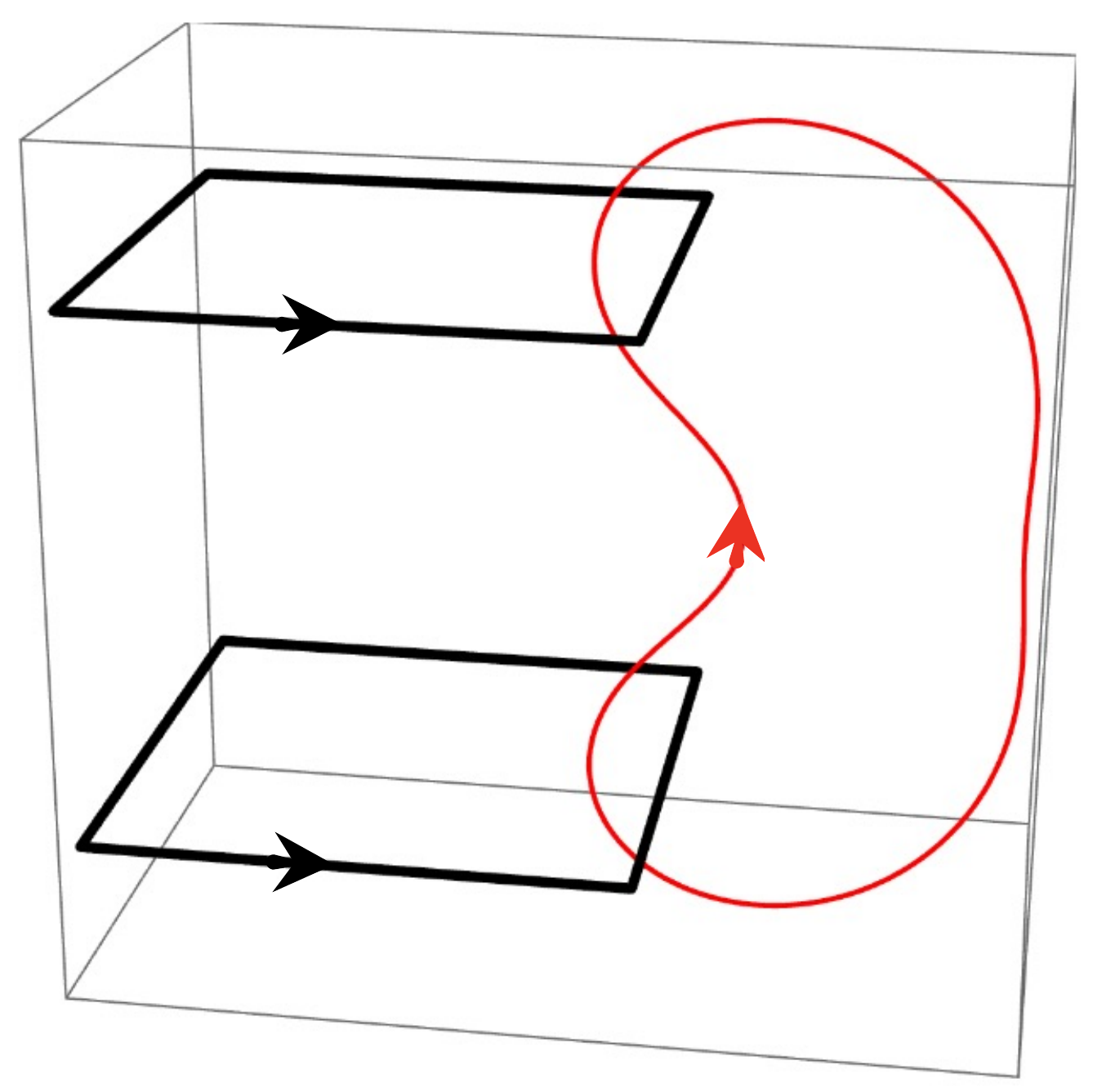}}
\vskip0in
\caption{Typical field line for identical square currents.}
\end{figure}

\vskip0in
\begin{figure}[h]
\hskip0in\scalebox{.22}[.22]{\includegraphics{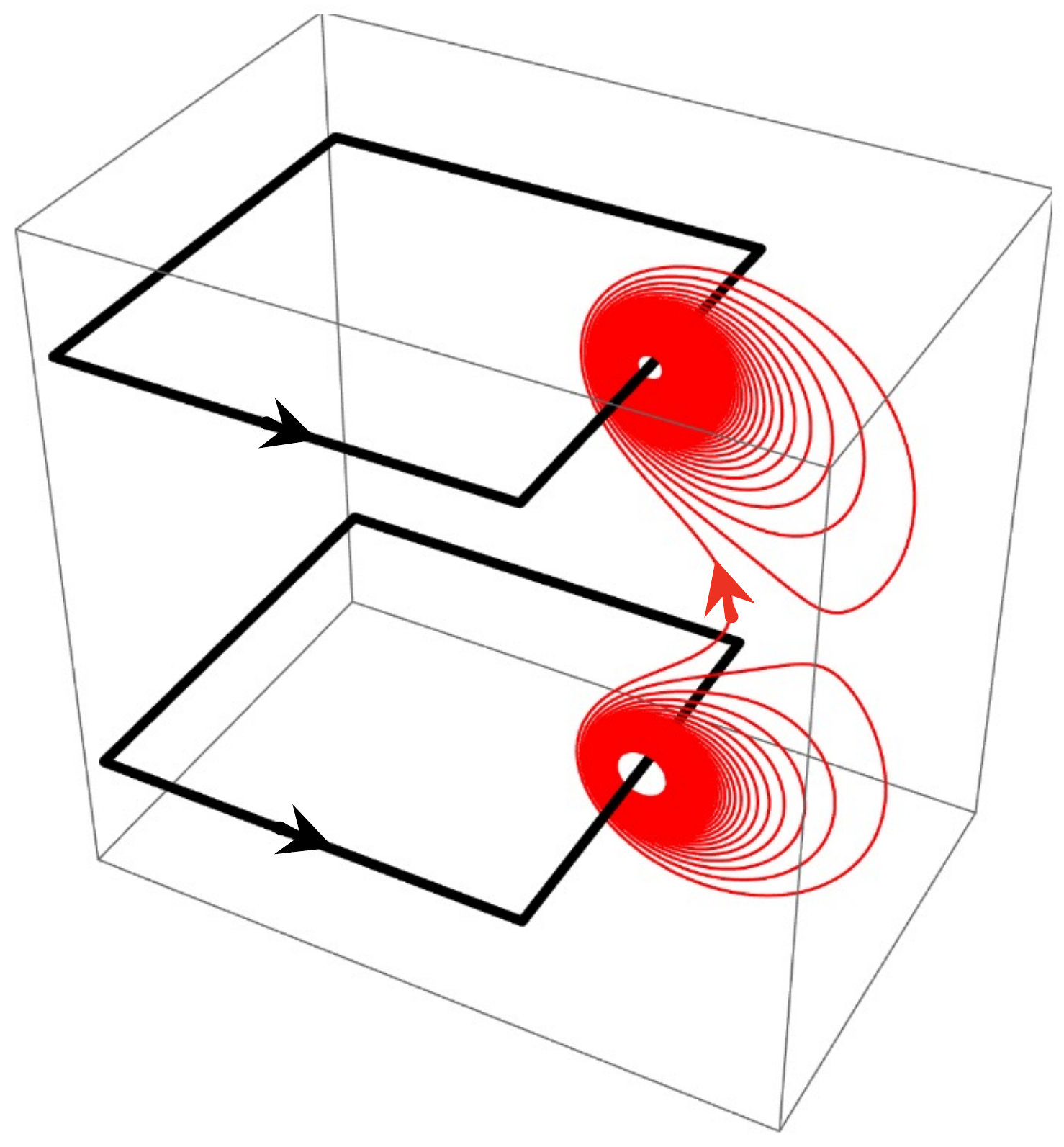}}
\vskip0in
\caption{Paired vortices for equal square currents.}
\end{figure}

It is also possible for a field line that begins as a vortex to fly off to infinity.  In Figure 9 we give the upper square twice the current and again start off in the plane bisecting one side (so the trajectory is confined to that plane).  In Figure 10 we do the same, but for the vertical plane including the back corners---again, symmetry dictates that the whole field line remains in that plane.  (Here the field line flies {\it in} from infinity.)

\vskip0in
\begin{figure}[h]
\hskip0.3in\scalebox{.23}[.23]{\includegraphics{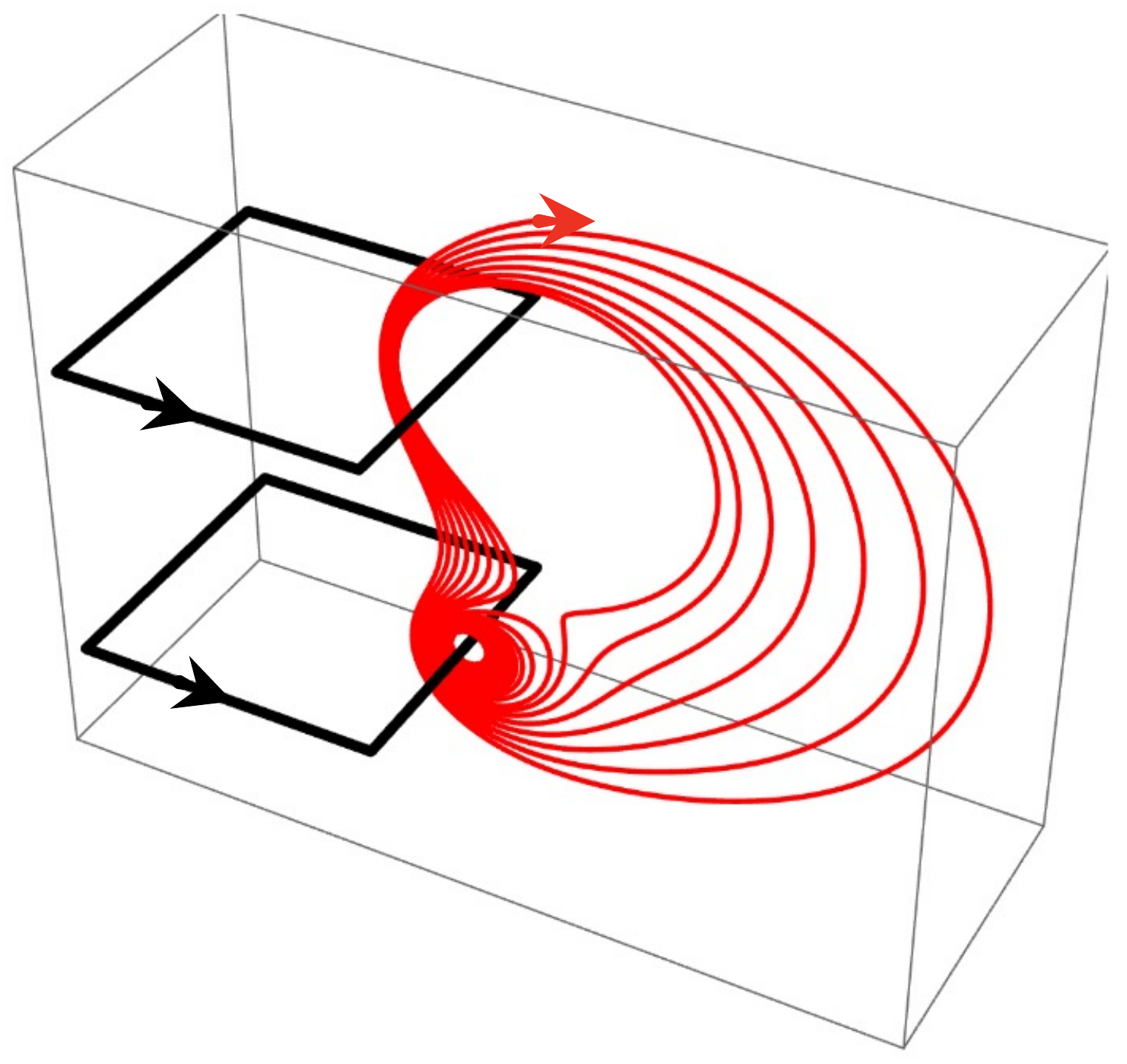}}
\vskip0in
\caption{Vortex to infinity for bisecting plane.}
\end{figure}

\vskip0in
\begin{figure}[h]
\hskip0in\scalebox{.22}[.22]{\includegraphics{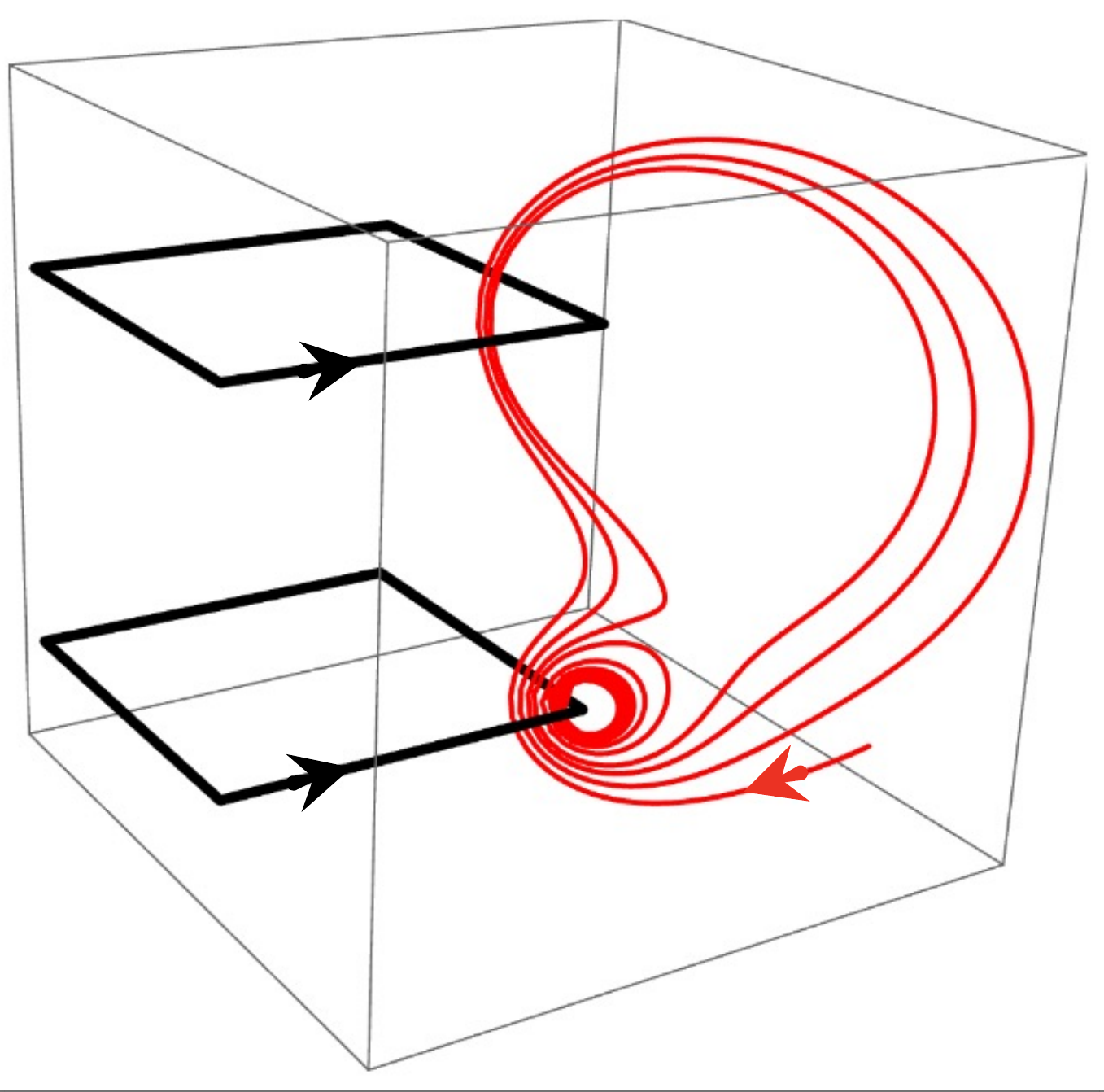}}
\vskip0in
\caption{Vortex to infinity for corner plane.}
\end{figure}
 
What if we start from a point that is neither in the bisecting plane nor in the corner plane?  Then symmetry no longer restricts the field line to a plane, and more complicated trajectories occur.  In the ``two spools" diagram (Fig.~11) the field line slinkies up the lower ring toward the corner, but instead of forming a vortex it opens out and is captured by the upper ring, where it slinkies down and transfers back to the first ring.  (For other initial conditions it escapes to infinity.)  The two symmetry planes constitute ``brick walls'' the field line cannot penetrate; between them it executes ``frustrated'' vortices and truncated slinkies.

\vskip0in
\begin{figure}[h]
\hskip0.1in\scalebox{.26}[.26]{\includegraphics{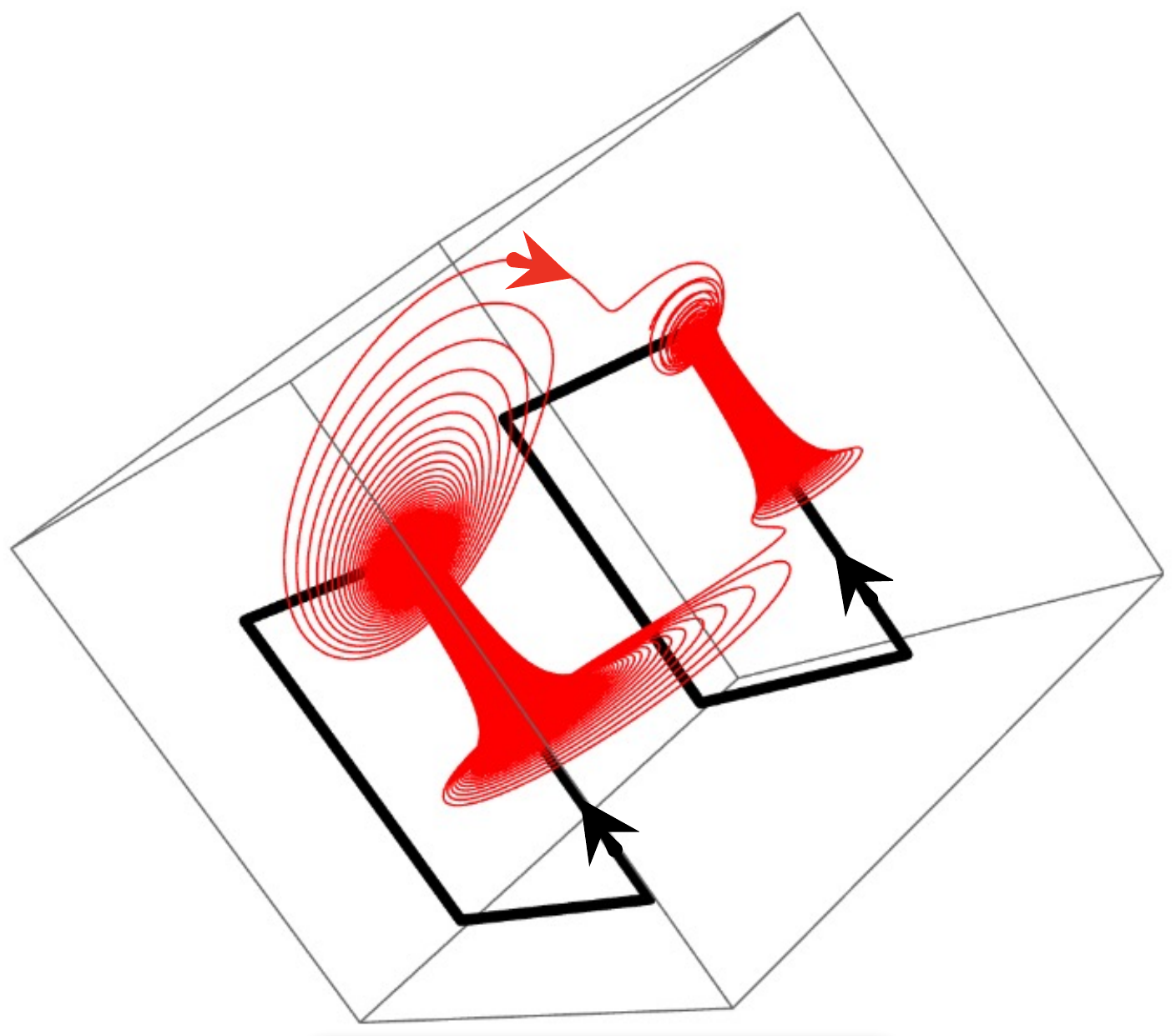}}
\vskip0in
\caption{Frustrated vortex and truncated slinky (same field as Figs.~9 and 10, but starting {\it between} the planes).}
\end{figure}

How is a vortex possible?  Close to a wire we certainly expect approximately circular field lines.  But here they are evidently perturbed by a field that aims {\it inward, toward the current}, for points in the plane of the ``circle'' (and since $\grad \cdot {\bf B}= 0$ this means that along the direction of the wire the perturbing field must point {\it away} from the center).  

Such a field occurs (for example) inside a spherical shell of radius $R$ carrying a surface current\cite{CASC}
\begin{equation}
{\bf K}(r,\theta,\phi) = K_0\sin(2\theta)\,\phat,
\end{equation}
where the field is
\begin{equation}
{\bf B}(s,\phi,z) = \frac{2\mu_0K_0}{5R}(-s\,\shat + 2z\,\zhat)
\end{equation}
(in cylindrical coordinates).  As a simple model for a vortex, then, we imagine a steady current $I$ running down the $z$ axis, passing through this spherical shell (centered at the origin).   In the $z=0$ plane the perturbing field aims inward, forcing the otherwise circular field line to spiral toward the axis (Fig.~12), while, for points above (or below) the plane the $z$ component produces a slinky riding up (or down) the $z$ axis (Fig.~13). 
\vskip0in
\begin{figure}[h]
\hskip0.2in\scalebox{.23}[.23]{\includegraphics{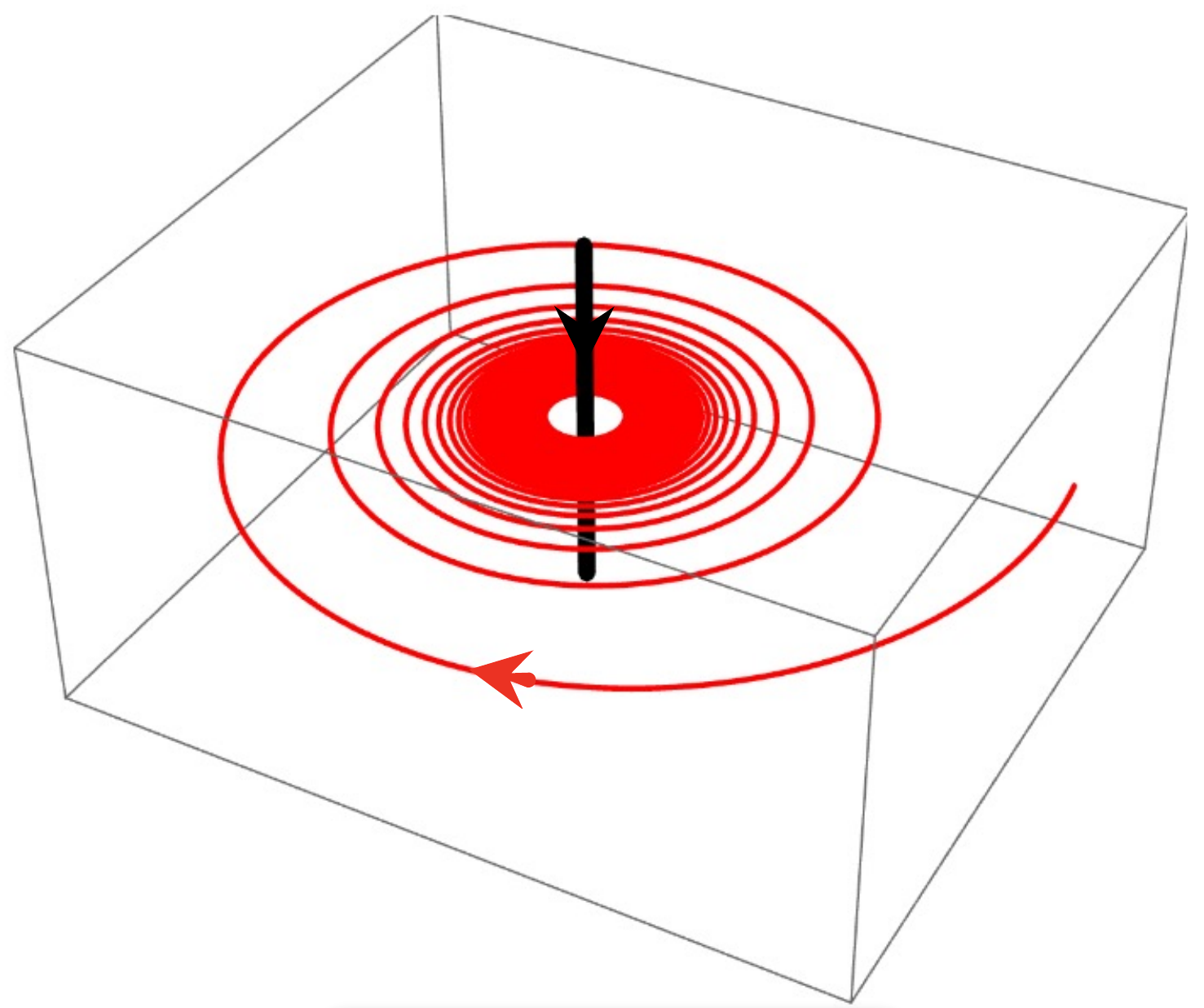}}
\vskip-.1in
\caption{Model vortex.  (Not shown: spherical surface current.)}
\end{figure}

\vskip0in
\begin{figure}[h]
\hskip0.2in\scalebox{.22}[.22]{\includegraphics{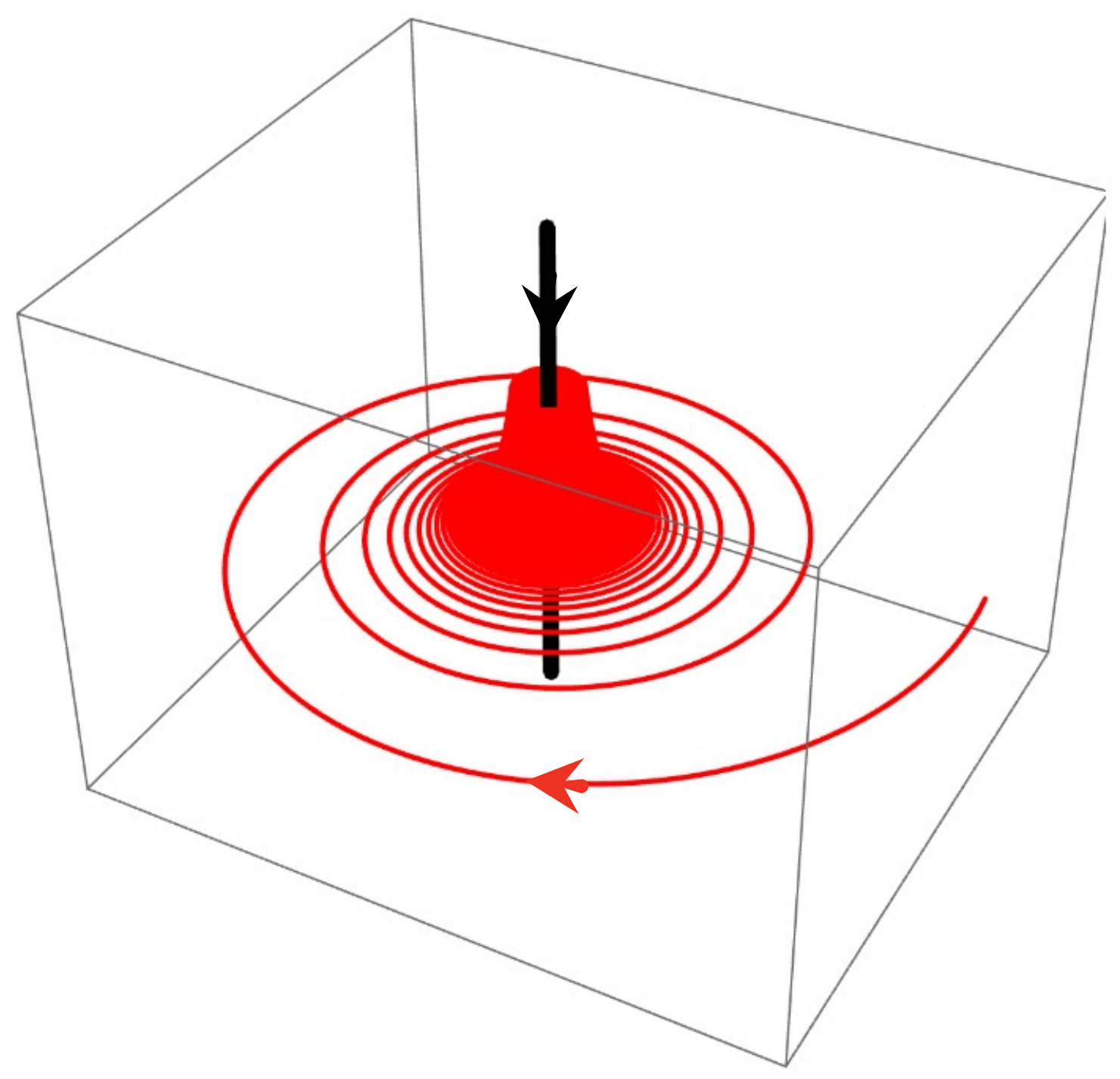}}
\vskip-.1in
\caption{A field line above the $x\,y$ plane.  (Not shown: spherical surface current.)}
\end{figure}

We can make this quantitative.  Equation \ref{eq2.6} yields the field line, in parametric form: 
\begin{equation}
s(u) =s_0e^{-\alpha u}, \ \phi(u) = \frac{\beta}{2\alpha s_0^2}\left(e^{2\alpha u}-1\right),\ z(u) = z_0e^{2\alpha u},
\end{equation}
where 
\begin{equation}
\alpha \equiv \frac{2\mu_0K_0}{5R}, \quad \beta \equiv \frac{\mu_0I}{2\pi},
\end{equation}
$s_0$ is the initial distance from the axis, $z_0$ is the initial distance above the $x\,y$ plane, and we set $\phi(0)=0$.  Eliminating $u$ in favor of $\phi$ as the independent variable, the field line takes the form
\begin{equation}
s(\phi) = \frac{s_0}{\sqrt{1+\gamma\phi}},\quad z(\phi) = z_0(1+\gamma \phi),
\end{equation}
with $\gamma\equiv (8\pi K_0s_0^2)/(IR)$.

\subsection{Chaotic Field Lines}

Over the years several authors\cite{RYSA} have explored magnetic fields that are ``chaotic,'' in the sense that field lines starting out very close together diverge exponentially as they progress.\cite{EATP}  Let us begin with two infinite perpendicular straight line currents, say one along the $z$ axis and one (with the same current) along the $y$ axis.  Near to an axis and far from the origin the field lines are of course approximately circular.  Closer to the origin a typical field line is a closed (but non-planar) loop (Fig.~14).    

\vskip0in
\begin{figure}[ht]
\hskip0.1in\scalebox{.24}[.24]{\includegraphics{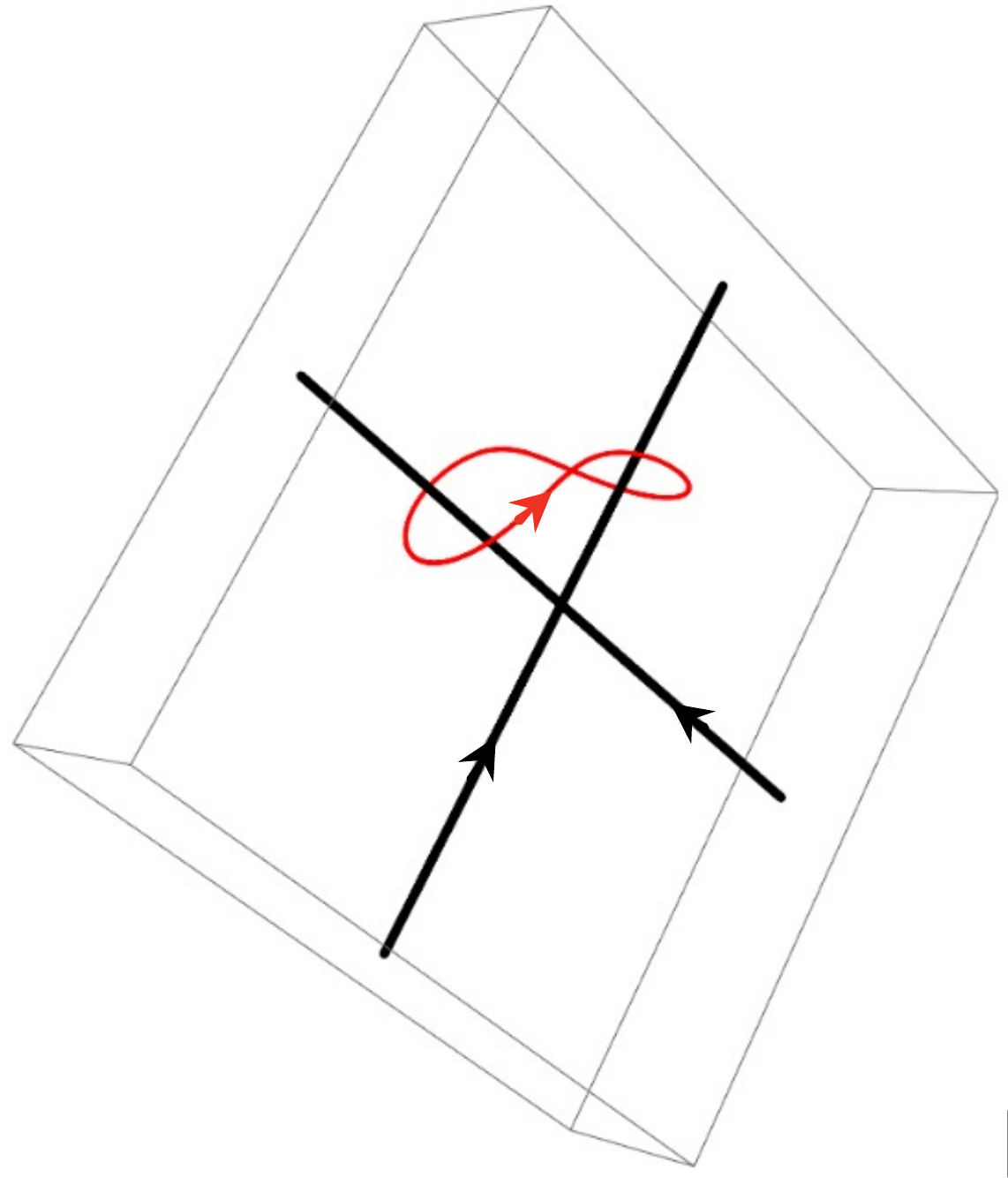}}
\vskip0in
\caption{Non-planar closed field line for intersecting perpendicular currents.}
\end{figure}

But suppose we move the second wire so that it is parallel to the $y$ axis, but at $x=b$ (and $z=0$).  This is one of the very first problems assigned to an early electronic computer.\cite{AEEC}  Figure 15 shows a typical field line.\cite{ATFL}

\vskip0in
\begin{figure}[ht]
\hskip0.1in\scalebox{.22}[.22]{\includegraphics{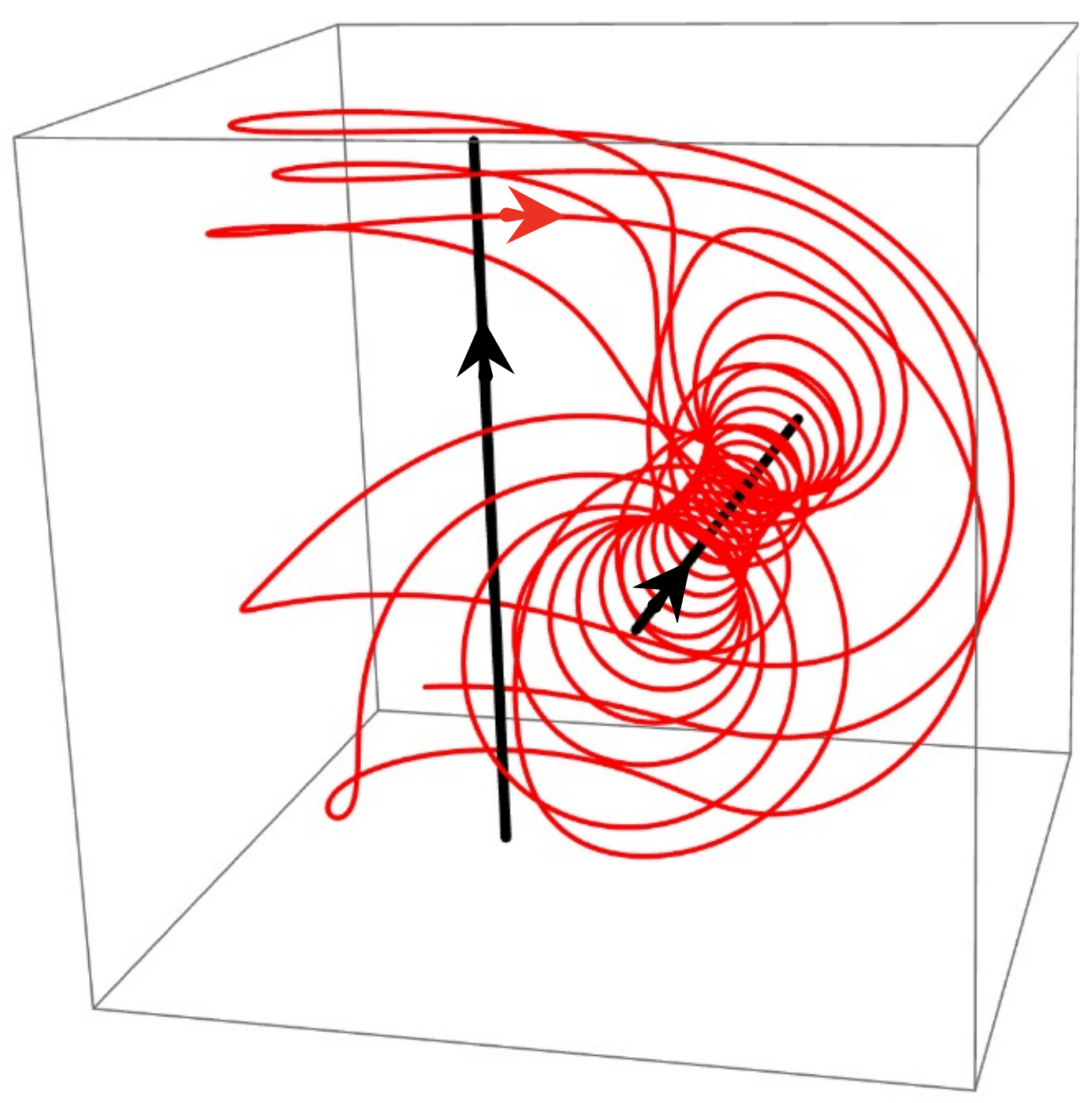}}
\vskip0in
\caption{``Chaotic'' field line for non-intersecting perpendicular currents.}
\end{figure}

\section{Symmetry}
The closure and planarity of field lines is closely associated with symmetries of the currents that produce them.  In this section we explore some examples.\cite{ESIC}

\subsection{Translational Symmetry}
By translational symmetry we mean that the current distribution is independent of one of the Cartesian coordinates---say, $z$.  We consider two cases: ``longitudinal'' (when the currents themselves are in the $z$ direction), and ``transverse'' (when the currents are perpendicular to the $z$ direction).  What are the resulting symmetries of {\bf B}?

\begin{enumerate}
\item {\bf Longitudinal Currents.}  Suppose
\begin{equation}
{\bf J}(x,y,z) = J_z(x,y)\,\zhat.\label{eq3}
\end{equation}
Then by the Biot-Savart law, 
\begin{equation}
{\bf B}({\bf r}) = \frac{\mu_0}{4\pi}\int \frac{{\bf J}({\bf r}')\times \brcurs}{\rcurs^3}\,d^3 {\bf r}'\label{eq4}
\end{equation}
(where $\brcurs \equiv {\bf r} - {\bf r}'$), the magnetic field takes the form\cite{FTTF}
\begin{equation}
{\bf B}(x,y,z) = B_x(x,y)\,\xhat + B_y(x,y)\,\yhat.\label{eq5}
\end{equation}
The field, and hence also the field {\it lines}, lie in planes of constant $z$---the same in every such plane.  (Example: an infinite straight wire.)

\item {\bf Transverse Currents.}  Suppose
\begin{equation}
{\bf J}(x,y,z) = J_x(x,y)\,\xhat + J_y(x,y)\,\yhat.\label{eq6}
\end{equation}
Then
\begin{equation}
{\bf B}(x,y,z) = B_z(x,y)\,\zhat.\label{eq7}
\end{equation}
The field lines are infinite, straight, and parallel to the $z$ axis.  This generalizes the familiar rule\cite{THEN}  for infinite solenoids of arbitrary cross-section: the field is parallel to the axis (and equal to $\mu_0 K$, where $K$ is the surface current density, for points inside).  
\end{enumerate}

Notice that Eq.~\ref{eq3} is to Eq.~\ref{eq5} as Eq.~\ref{eq7} is to Eq.~\ref{eq6}, illustrating a kind of duality: {\bf B} is to {\bf J} as {\bf J} is to $(\nabla^2) {\bf B}$.  (This follows from $\grad\cdot {\bf B} = 0$, $\grad\cdot {\bf J} = 0$, and $\grad\times{\bf B} = \mu_0 {\bf J}$.)

\subsection {Azimuthal Symmetry}
By azimuthal symmetry we mean that the current is independent of $\phi$.  Again, we consider two cases:

\begin{enumerate}
\item {\bf Toroidal Currents.}
Suppose 
\begin{equation}
{\bf J}(s,\phi,z) = J_s(s,z)\,\shat + J_z(s,z)\,\zhat,\label{eq8}
\end{equation}
in cylindrical coordinates.  Then 
\begin{equation}
{\bf B}(s,\phi,z) = B_\phi(s,z)\,\phat;\label{eq9}
\end{equation}
the field points in the azimuthal direction,\cite{ITAD} and the field lines are closed (planar) circles.  This includes as a special case all azimuthally symmetric {\bf longitudinal currents}:
\begin{equation}
{\bf J}(s,\phi,z) = J_z(s)\,\zhat,\quad  {\bf B}(s,\phi,z) = B_\phi(s)\,\phat.\label{eq10}
\end{equation}
\item {\bf Azimuthal Currents.}  Suppose
\begin{equation}
{\bf J}(s,\phi,z) = J_\phi(s,z)\,\phat.\label{eq11}
\end{equation}
Then
\begin{equation}
{\bf B}(s,\phi,z) = B_s(s,z)\,\shat + B_z(s,z)\,\zhat.\label{eq12}
\end{equation}
The field lines lie in planes of constant $\phi$, but it is not clear whether they must be closed.  We'll investigate this further in Appendix B.
\end{enumerate}

Again, Eq.~\ref{eq8} is to Eq.~\ref{eq9} as Eq.~\ref{eq12} is to Eq.~\ref{eq11}, illustrating the duality between {\bf J} and {\bf B}.

\subsection{Mirror Symmetry}
Suppose the source current has no $z$ component
\begin{equation}
{\bf J}(x,y,z) = J_x(x,y,z)\,\xhat + J_y(x,y,z)\,\yhat\label{eq13}
\end{equation}
and is symmetric with respect to the $x\,y$ plane:
\begin{align}
J_x(x,y,-z) &= J_x(x,y,z),\nonumber\\
J_y(x,y,-z) &= J_y(x,y,z).\label{eq14}
\end{align}
Then the magnetic field satisfies
\begin{align}
B_x(x,y,-z) &= -B_x(x,y,z),\nonumber\\
B_y(x,y,-z) &= -B_y(x,y,z),\label{eq15}\\
B_z(x,y,-z) &=B_z(x,y,z).\nonumber
\end{align}
As always, this follows from the Biot-Savart law (see Appendix A).  We'll call the combination ``mirror'' symmetry.  Notice that the fields at $z=0$ point in the $z$ direction (Eq.~\ref{eq15}), and hence any field line that crosses the $x\,y$ plane does so perpendicularly.

{\bf Planar Currents.}  Suppose the current lies entirely in a plane (make it the $x\,y$ plane).  This is a special case of mirror symmetry (Eqs.~\ref{eq13} and \ref{eq14} hold trivially), so the fields satisfy Eq.~\ref{eq15}.  {\it Question:}  What field line configurations are possible?

Consider first a field line that never crosses the $x\,y$ plane.  It cannot form a closed loop (with a matching loop at the image position), because, by Amp\`ere's law, the line integral of {\bf B} around that loop would be 
\begin{equation}
\oint {\bf B}\cdot d{\bf r} = \int {\bf B} \cdot \frac{d{\bf r}}{du}\,du = \int B^2\,du = \mu_0I_{\rm enc} = 0.\label{eq16}
\end{equation}
So {\bf B} would have to be zero everywhere along the field line---which is no field line at all.  Nor could it form a slinky or a vortex---they wrap around currents, so it would have to cross the $x\,y$ plane.  But it could go to infinity---for instance, the field of an infinite sheet of current in, say, the $x$ direction.  Could it terminate in mid-air?  Remember, this can happen at a point where ${\bf B}={\bf 0}$.  Consider two concentric circular loops carrying equal but opposite currents, the outer one with twice the radius ($a$) of the inner one.  The field on the axis is
\begin{equation}
{\bf B}(z) = \frac{\mu_0Ia^2}{2}\left[\frac{4}{(4a^2+z^2)^{3/2}}-\frac{1}{(a^2+z^2)^{3/2}}\right]\zhat,\label{eq17}
\end{equation}
and it  goes through zero at $z_0 \approx  0.9869\,a$.  If $I$ is positive, the field points in the negative $z$ direction for $-z_0 < z < z_0$, and in the positive $z$ direction otherwise; there is a field line that starts at $(0,0,z_0)$ and runs up the $z$ axis, and another that starts there and runs down to $-z_0$ (but that one crosses the $x\,y$ plane, of course), and a third that starts at $-\infty$ and runs up the $(-)z$ axis to $-z_0$.  (There are also ``horizontal'' field lines that converge on $z_0$ and diverge from $-z_0$.)  So the answer is ``yes'': such field lines can terminate in midair.

What about field lines that {\it do} cross the $x\,y$ plane?  If they cross just once, they can either terminate in mid-air (we just saw an example), or they can run off to infinity (for instance, on the axis of a circular current loop).  If they cross twice, then necessarily they form closed loops---one lobe above the plane joining its mirror image below the plane.  They cannot cross multiple times (forming, say, a slinky or a vortex), for such configurations by their nature violate mirror symmetry.  

{\it Conclusion:} The field lines generated by planar currents either form closed loops, or they run off to infinity, or they terminate in mid-air.\cite{TTIM}

\section{Conclusion}
Magnetic field lines can form finite closed loops, they can escape to infinity, they can terminate in midair at points where the field is zero, they can end in death spirals at a line current, or they can wander around forever as a slinky that never closes; they can even form a chaotic rat's nest.  Notice that every figure in this paper, save the first and the last, shows a {\it single} field line!  (So  much for the notion that the density of field lines reflects the strength of the field.)  What, then, is the answer to our original question:  Why do the familiar steady current configurations produce closed magnetic field lines?  There doesn't seem to be a simple generic answer; it all depends on the symmetry of the current.

\begin{itemize}
\item {\bf Infinite straight current.}  A steady current $I$ runs along the $z$ axis.  This is an example of a longitudinal current with azimuthal symmetry (Eq.~\ref{eq10}); the field is azimuthal, and the field lines are (coaxial) closed circles.
\item {\bf Toroidal coil.}  A surface current ${\bf K}(s,z) = K_s(s,z)\,\shat + K_z(s,z)\,\zhat$ flows over a toroid about the $z$ axis (perhaps with circular or rectangular cross-section, but it doesn't matter, as long as it is uniform all the way around).  This is an example of a toroidal current with azimuthal symmetry (Eq.~\ref{eq8}), and the field lines are (coaxial) circles.  (There are no field lines exterior to the toroid, where the field is zero.)
\item {\bf Circular current loop.}  A current $I$ flows in a circle that lies in the $x\,y$ plane, centered at the origin.  This is an example of an azimuthal current with azimuthal symmetry (Eq.~\ref{eq11}), and the field lines lie in planes of constant $\phi$ (Eq.~\ref{eq12}).  The current is planar, so the field lines are closed (except along the axis, where they run off to infinity).  (The same goes for an ideal (point) magnetic dipole, which is the limiting case of a circular current loop.)
\item {\bf Spinning figures of revolution.} Other familiar examples include uniformly charged spinning spheres and finite circular solenoids (or objects with equivalent currents: uniformly magnetized balls and cylindrical bar magnets).  These are again examples of azimuthal currents with azimuthal symmetry (Eq.~\ref{eq11}), so the field lines lie in planes at constant $\phi$ (Eq.~\ref{eq12}).  We could regard them as a stack of coaxial circular current loops of varying radius.  The simplest example would be {\it two} identical rings, at $z=\pm b$.  Typical field lines circle one loop (as in Fig.~2), or both (as in Fig.~7), but could we get vortices (as in Fig.~8)?  No we cannot: a vortex would require a field component pointing {\it inward} (toward the wire), in the constant $\phi$ plane, and therefore pointing {\it away} from the center of the spiral (i.e. in the $\pm \phat$ directions) for points perpendicular to that plane.  But a field in the $\phat$ direction is excluded by the azimuthal symmetry.  Evidently these field lines, too, must form closed loops (or, along the axis, run off to infinity).\cite{ROTI} 
\end{itemize}

Yes: simple systems (straight line currents, circular loops, tightly wound toroidal coils, bar magnets, \ldots) produce closed magnetic field lines.  But the full story is so much richer!

\bigskip

\centerline{\bf {Appendix A: Proof of Equation 20}}

\bigskip

If the current has mirror symmetry (Eqs.~\ref{eq13} and \ref{eq14}), then 
\begin{align}
{\bf J}({\bf r}')\times \brcurs& = \begin{vmatrix} \xhat&\yhat&\zhat\\J_x&J_y&0\\ \rcurs_x&\rcurs_y&\rcurs_z\end{vmatrix}\label{eq34}\\
=&(J_y\rcurs_z)\,\xhat - (J_x\rcurs_z)\,\yhat + (J_x\rcurs_y-J_y\rcurs_x)\,\zhat,\nonumber
\end{align}
so (letting $z'' = -z'$), the Biot-Savart law (Eq.~\ref{eq4}) says
\begin{align}
&B_x(x,y,z)\label{eq35}\\
&=\frac{\mu_0}{4\pi}\int\frac{J_y(x',y',z')(z-z')}{[(x-x')^2+(y-y')^2+(z-z')^2]^{3/2}}\,dx'\,dy'\,dz'\nonumber\\
&=\frac{\mu_0}{4\pi}\int\frac{J_y(x',y',-z'')(z+z'')}{[(x-x')^2+(y-y')^2+(z+z'')^2]^{3/2}}\,dx'dy'dz''\nonumber\\
&=\frac{\mu_0}{4\pi}\int\frac{J_y(x',y',z'')(z+z'')}{[(x-x')^2+(y-y')^2+(z+z'')^2]^{3/2}}\,dx'\,dy'\,dz''\nonumber
\end{align}
(we used Eq.~\ref{eq14} in the last line).  Thus
\begin{align}
&B_x(x,y,-z)\nonumber\\
&=\frac{\mu_0}{4\pi}\int\frac{J_y(x',y',z'')(-z+z'')}{[(x-x')^2+(y-y')^2+(-z+z'')^2]^{3/2}}\,dx'\,dy'\,dz''\nonumber\\
&=-B_x(x,y,z),\label{eq36}
\end{align}
and the same goes for $B_y$.  That establishes the first two equations in Eq.~\ref{eq15}.  Meanwhile, 
\begin{align}
&B_z(x,y,z)\nonumber\\
&=\frac{\mu_0}{4\pi}\int\frac{J_x(x',y',z')(y-y')- J_y(x',y',z')(x-x')}{[(x-x')^2+(y-y')^2+(z-z')^2]^{3/2}}\nonumber\\&\qquad\qquad\,dx'\,dy'\,dz'\nonumber\\
&=\frac{\mu_0}{4\pi}\int\frac{J_x(x',y',-z'')(y-y')- J_y(x',y',-z'')(x-x')}{[(x-x')^2+(y-y')^2+(z+z'')^2]^{3/2}}\nonumber\\
&\qquad\qquad \,dx'dy'dz''\nonumber\\
&=\frac{\mu_0}{4\pi}\int\frac{J_x(x',y',z'')(y-y')- J_y(x',y',z'')(x-x')}{[(x-x')^2+(y-y')^2+(z+z'')^2]^{3/2}}\nonumber\\
&\qquad\qquad\,dx'\,dy'\,dz''\nonumber\\
&=B_z(x,y,-z),\label{eq37}
\end{align}
which confirms the third equation in Eq.~\ref{eq15}.

\bigskip

\centerline{\bf Appendix B: Field Lines and Contour Plots}

\bigskip

In this Appendix we explore further the case of azimuthal currents with azimuthal symmetry (Eq.~\ref{eq11}):
\begin{equation}
{\bf J}(s,\phi,z) = J_\phi(s,z)\,\phat.\label{eqB27}
\end{equation}
The magnetic field takes the form (Eq.~\ref{eq12})
\begin{equation}
{\bf B}(s,\phi, z) = B_s(s,z)\,\shat + B_z(s,z)\, \zhat,\label{eqB28}
\end{equation}
and its components satisfy
\begin{equation}
\grad \cdot {\bf B} = \frac{1}{s}\frac{\partial}{\partial s}(sB_s) + \frac{\partial B_z}{\partial z} = 0.\label{eqB29}
\end{equation}
The field lines lie in planes at constant $\phi$, and numerical plots are strikingly reminiscent of contour maps.  Does there exist a scalar function $U(s,z)$ such that the contours of $U$ correspond to the magnetic field lines?  That would require that {\bf B} be perpendicular to the gradient of $U$:
\begin{equation}
{\bf B}\cdot \grad U = B_s\frac{\partial U}{\partial s} + B_z\frac{\partial U}{\partial z} = 0.\label{eqB30}
\end{equation}
We can satisfy Eq.~\ref{eqB30}, and automatically also Eq.~\ref{eqB29}, if we choose
\begin{equation}
\frac{\partial U}{\partial s} = sB_z,\quad \frac{\partial U}{\partial z} = -sB_s.\label{eqB31}
\end{equation}  

For the current in Eq.~\ref{eqB27} the vector potential is
\begin{equation}
{\bf A}({\bf r}) = \frac{\mu_0}{4\pi} \int\frac{{\bf J}({\bf r'})}{\rcurs}\,d^3{\bf r}' = \frac{\mu_0}{4\pi}\int\frac{J_\phi(s',z')\,\phat'}{|{\bf r}-{\bf r}'|}\,d^3{\bf r}'.\label{eqB32}
\end{equation}
We might as well choose axes such that ${\bf r}$ lies in the $x\,z$ plane, at $y=0$.  An element of current at $P'$ is matched by an element at $P''$ with the same $s'$ and $z'$, but opposite $\phi'$ (which we can run from $-\pi$ to $+\pi$, instead of $0$ to $2\pi$); $P'$ and $P''$ share the same $\rcurs=|{\bf r}-{\bf r}'|$, but the currents point in the directions $\phat'$ and $\phat''$, and when the two vectors are added, the resultant {\it points in the $y$ direction}.  {\it Conclusion:}  The vector potential, in the $x\,z$ plane, points purely in the $y$ direction---which is to say, in general, that it points in the $\phat$ direction:
\begin{equation}
{\bf A}({\bf r}) = \frac{\mu_0}{4\pi}\,\phat \int\frac{J_\phi(s',z')}{|{\bf r}-{\bf r}'|}\,s'\,ds'\,d\phi'\,dz'.\label{eqB33}
\end{equation}
Now
\begin{equation}
\brcurs = (s\cos\phi - s'\cos\phi')\xhat + (s\sin\phi - s'\sin\phi')\yhat  + (z-z')\zhat,\label{eqB34}
\end{equation}
so
\begin{equation}
\rcurs^2 = s^2 + (s')^2 + (z-z')^2-2ss'(\cos\phi \cos\phi' + \sin\phi \sin\phi'),\label{eqB35}
\end{equation}
and the $\phi'$ integral becomes
\begin{align}
I_\phi &= \int_0^{2\pi} \frac{d\phi'}{\sqrt{q^2 - 2ss'\cos(\phi-\phi')}} \nonumber\\
&= \int_0^{2\pi} \frac{d\phi''}{\sqrt{q^2 - 2ss'\cos(\phi'')}}  ,\label{eqB36}
\end{align}
where $q^2\equiv s^2+(s')^2 + (z-z')^2$.  We could actually {\it do} this integral, but we don't need it \ldots\ the point is that the result is {\it independent of $\phi$}, and hence
\begin{equation}
{\bf A}({\bf r}) = A_\phi(s,z)\,\phat.\label{eqB37}
\end{equation}
Of course, $\grad \times {\bf A} = {\bf B}$, or
$$-\frac{\partial A_\phi}{\partial z} = B_s, \quad \frac{1}{s}\frac{\partial}{\partial s}(sA_\phi) = B_z,$$
so the contour function $U$ (Eq.~\ref{eqB31}) is precisely
\begin{equation}
U(s,z) = s\,A_\phi(s,z).\label{eqB38}
\end{equation}

The field lines are contours of the function $U(s,z)$, and they are typically closed loops (for this symmetry) just as contour lines are typically closed loops.  Not always, however: imagine a ridge at constant altitude---perhaps even forming a spiral---the contour along the top does not form a closed loop.  And this correspondence works only for two-dimensional configurations; a slinky is nobody's contour line.

{\bf Example:}  Consider a charged spherical shell (radius $R$, charge density $\sigma$), spinning around the $z$ axis with angular velocity $\omega$ (ref.~19, Example 5.11).  The vector potential is
\begin{equation}
{\bf A} = \begin{cases}\displaystyle ks\,\phat,& (r\leq R),\\
\displaystyle k\frac{R^3s}{r^3}\,\phat,&(r\geq R),\label{eqB39}
\end{cases}
\end{equation}
where $k \equiv(\mu_0 \omega\sigma R)/3$.  So 
\begin{equation}
U(s,z) = \begin{cases}\displaystyle ks^2,& (\sqrt{s^2+z^2}\leq R),\\
\displaystyle \frac{kR^3s^2}{(s^2+z^2)^{3/2}},&(\sqrt{s^2+z^2}\geq R).\label{eqB40}
\end{cases}
\end{equation}
The contour plot is shown in Fig.~16.

\vskip0in
\begin{figure}[h]
\hskip-.2in\scalebox{.4}[.4]{\includegraphics{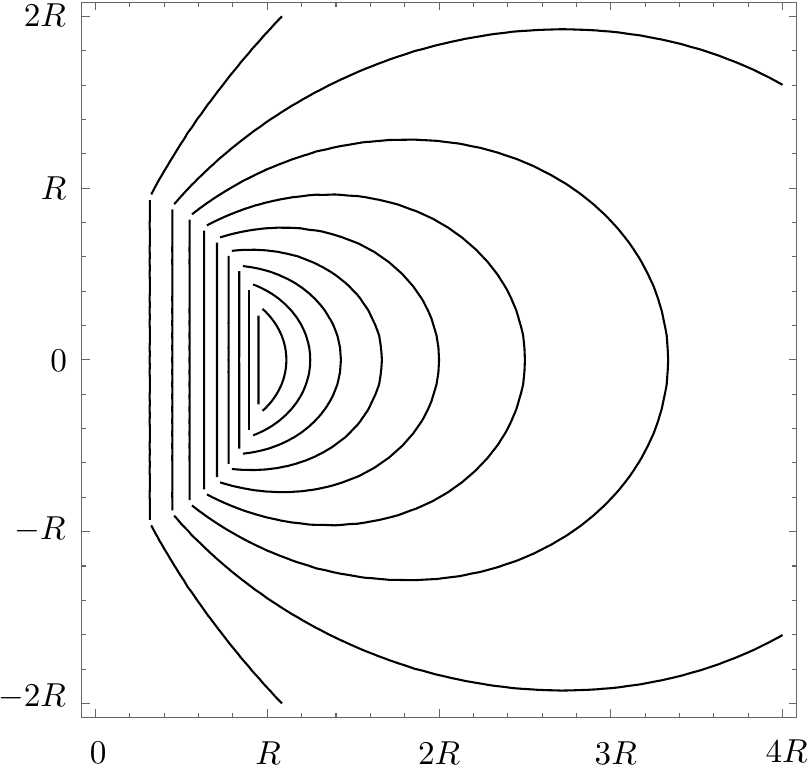}}
\vskip0in
\caption{Contour plot for Eq.~\ref{eqB40}.}
\end{figure}

\bigskip
\bigskip

\centerline{\bf Appendix C: Curvature and Torsion}

\bigskip

If we use arc length for the parameter $u$, then
\begin{equation}
\that \equiv \frac{d{\bf r}}{du},\label{eqC41}
\end{equation}
is a unit tangent vector along the field line.  Its derivative defines the curvature $\kappa$ (which measures the departure from straightness) and the unit vector $\nhat$:\cite{ATUV}
\begin{equation}
\frac{d\that}{du} = \kappa\,\nhat;\label{eqC42}
\end{equation}
$\nhat$ is perpendicular to $\hat {\bf t}$, and their cross-product defines a third unit vector,
\begin{equation}
\bhat \equiv \that \times \nhat.\label{eqC43}
\end{equation}
At any given point along the curve, $\that$, $\nhat$, and $\bhat$ form a (right-handed) triad of orthogonal unit vectors.  Their derivatives satisfy the Frenet-Serret formulas:\cite{TFSF}
\begin{equation}
\frac{d\that}{du} = \kappa\,\nhat,\quad \frac{d\nhat}{du} = \tau\, \bhat -\kappa \, \that,\quad \frac{d\bhat}{du} = -\tau \,\nhat.\label{eqC44}
\end{equation}
Here $\tau$ is the torsion; it measures the departure from flatness.  A planar curve (with nonzero $\kappa$) has $\tau = 0$ everywhere.

In the case of magnetic field lines, $\that$ is given by Eq.~\ref{eq2.5}:
\begin{equation}
\that(u) = \frac{d{\bf r}}{du} = \hat{\bf B}\left({\bf r}(u)\right),\label{eqC45}
\end{equation}
where $\hat{\bf B}$ is a unit vector in the direction of the field.  Differentiating the $i$th component with respect to $u$,
\begin{equation}
\left(\frac {d\that}{du}\right)_i = \frac{\partial\hat{\bf B}_i}{\partial{\bf r}_j}\frac{d{\bf r}_j}{du} = \left(\grad_j \hat{\bf B}_i\right)\hat{\bf B}_j = (\hat {\bf B} \cdot \grad)\,\hat {\bf B}_i,\label{eqC46}
\end{equation}
(summation over $j$ implied), so
\begin{equation}
\kappa\,\nhat = (\hat{\bf B}\cdot \grad)\hat {\bf B} = -\hat{\bf B} \times(\grad \times \hat{\bf B}).\label{eqC47}
\end{equation}
(The final step follows from the product rule for\linebreak $\grad({\bf A}\cdot {\bf B})$.)  The other derivatives in Eq.~\ref{eqC44} can be handled in the same way.

{\bf Example:}  Take the case of an infinite wire carrying a steady current $I$.  The field, in cylindrical coordinates $(s, \phi, z)$, is
\begin{equation}
{\bf B}= \frac{\mu_0I}{2\pi s}\,\phat\quad \Rightarrow\quad \that = \hat{\bf B} = \phat.\label{eqC48}
\end{equation}
From Eq.~\ref{eqC47},
\begin{equation}
\kappa\,\nhat =  -\phat \times(\grad\times \phat) =-\phat \times\left(\frac{1}{s}\,\zhat\right)  = -\frac{1}{s}\,\shat,\label{eqC49}
\end{equation}
so 
\begin{equation}
\kappa = \frac{1}{s},\quad \nhat = -\shat.\label{eqC50}
\end{equation}
This makes sense: the field lines are circles of radius $s$ (curvature $1/s$), and $\nhat$ is a unit vector pointing toward the center of the circle (on the $z$ axis).  Meanwhile
\begin{equation}
\bhat = \that \times \nhat = \phat \times (-\shat) = \zhat,\label{eqC51}
\end{equation}
and $(\that, \nhat,\bhat) = (\phat, -\shat, \zhat)$ constitute a right-handed triplet of orthogonal unit vectors, as promised.  Finally, 
\begin{equation}
\frac{d\nhat}{du} = -\frac{1}{s}\,\phat = -\kappa\,\that,\label{eqC52}
\end{equation}
and hence (from Eq.~\ref{eqC44}), the torsion $\tau = 0$.  This too makes sense: the field lines are planar (circles).  Evidently $d\bhat/du=(\that\cdot \grad)\bhat={\bf 0}$, and this is indeed the case.

Suppose we now introduce a uniform magnetic field in the $z$ direction, so the field lines are helices (Fig.5):
\begin{equation}
{\bf B} = \frac{\mu_0I}{2\pi s}\,\phat + B_0\,\zhat = \frac{\mu_0 I}{2\pi s}(\phat + \alpha s\,\zhat)\label{eq30}
\end{equation}
(where $\alpha \equiv 2\pi B_0/\mu_0I$).  Then
\begin{equation}
\that = \hat{\bf B} = \frac{\phat + \alpha s\,\zhat}{\sqrt{1+\alpha^2s^2}}.\label{eqC53}
\end{equation}
From Eqs.~\ref{eqC47} and \ref{eqC43} we get
\begin{equation}
\kappa = \frac{1}{s(1+\alpha^2s^2)}, \ \nhat = -\shat,\  {\rm and}\  \bhat = \frac{-\alpha s\,\phat +\zhat}{\sqrt{1+\alpha^2s^2}},\label{eqC55}
\end{equation}
while Eq.~\ref{eqC44} yields the torsion
\begin{equation}
\tau = \frac{\alpha}{1+\alpha^2s^2}\label{eqC56}
\end{equation}

\section*{Acknowledgments}  
We thank the anonymous referees for their careful reading of the manuscript, for helpful comments, and especially for calling our attention to several pertinent references.


\begin{thebibliography}{99}
\bibitem{AFL} Magnetic field lines were introduced by Michael Faraday, who was inspired by the pattern of iron filings in the vicinity of a magnet.  They are sometimes called ``magnetic flux lines,'' or even ``magnetic lines of force'' (which is grossly misleading).
\bibitem{JIME} In highly conductive plasmas (such as the sun's corona) the special environment afforded by magnetohydrodynamics endows magnetic field lines with a kind of physical ``reality'' that they do not enjoy more generally.  They can be spectacularly visible in photographs of solar flares, for instance (but of course we are not really seeing the field lines themselves---only radiation from the charged particles that track them).
\bibitem{CASC} It might be a spinning sphere with surface charge $\sigma(\theta) = \sigma_0\cos\theta$ and angular velocity $\omega_0$ (about the $z$ axis).  In that case $K_0 = \sigma_0\omega_0R/2$.
\bibitem{EATP} It is not always clear from simulations whether field lines that start out chaotic remain so indefinitely, as we follow them out; Hosoda, ref.~(14), refers to this as ``transient chaos.''
\bibitem{ATFL} Lieberherr, ref.~(11), recommends adding a small constant magnetic field, to prevent (or at least delay) the field line's tendency to run off to infinity.  M.~Hosoda, ref.~(14) does the same thing for a different reason, noting that in practice such a perturbation is typically present in the form of the earth's magnetic field.
\bibitem{FTTF} The calculation is essentially the same for each symmetry, so we will do it just once, in Appendix A.
\bibitem{ATUV} We take $\kappa = 1/R$ (where $R$ is the radius of the osculating circle) to be non-negative; then $\nhat$ points toward the center of that circle.
\end{thebibliography}
\end{document}